\documentclass[prb,twocolumn,showpacs,amsmath,amssymb,superscriptaddress,longbibliography]{revtex4-1}
\usepackage{dcolumn}% Align table columns on decimal point
\usepackage{bm,graphicx}% bold math
\usepackage{etoolbox}
\apptocmd{\thebibliography}{\raggedright}{}{}
\usepackage{url}
\usepackage[colorlinks]{hyperref}
\usepackage{breakurl}

\begin{document}

\title{Strong interband Faraday rotation in 3D topological insulator Bi$_2$Se$_3$}
\author{L. Ohnoutek}
\affiliation{Institute of Physics, Charles University, Ke Karlovu 5, CZ-121 16 Praha 2, Czech Republic}
\author{M. Hakl}
\affiliation{Laboratoire National des Champs Magn\'etiques Intenses,
CNRS-UJF-UPS-INSA, 25, avenue des Martyrs, 38042 Grenoble, France}
\author{M. Veis}
\affiliation{Institute of Physics, Charles University, Ke Karlovu 5, CZ-121 16 Praha 2, Czech Republic}
\author{B. A. Piot}
\affiliation{Laboratoire National des Champs Magn\'etiques Intenses,
CNRS-UJF-UPS-INSA, 25, avenue des Martyrs, 38042 Grenoble, France}
\author{C. Faugeras}
\affiliation{Laboratoire National des Champs Magn\'etiques Intenses,
CNRS-UJF-UPS-INSA, 25, avenue des Martyrs, 38042 Grenoble, France}
\author{G. Martinez}
\affiliation{Laboratoire National des Champs Magn\'etiques Intenses,
CNRS-UJF-UPS-INSA, 25, avenue des Martyrs, 38042 Grenoble, France}
\author{M. V. Yakushev}
\affiliation{Department of Physics, SUPA, Strathclyde University, G4 0NG Glasgow, UK}
\affiliation{Ural Federal University and Institute of Solid State Chemistry of RAS, Ekaterinburg, 620002, Russia}
\author{R.~W.~Martin}
\affiliation{Department of Physics, SUPA, Strathclyde University, G4 0NG Glasgow, UK}
\author{\v{C}. Dra\v{s}ar}
\affiliation{Institute of Applied Physics and Mathematics, Faculty of Chemical Technology, University of Pardubice, Studentsk\'a 84, 532 10 Pardubice, Czech Republic}
\author{A. Materna}
\affiliation{Institute of Electronic Materials Technology, ul. Wolczynska 133, PL 01-919 Warsaw, Poland}
\author{G. Strzelecka}
\affiliation{Institute of Electronic Materials Technology, ul. Wolczynska 133, PL 01-919 Warsaw, Poland}
\author{A. Hruban}
\affiliation{Institute of Electronic Materials Technology, ul. Wolczynska 133, PL 01-919 Warsaw, Poland}
\author{M. Potemski}
\affiliation{Laboratoire National des Champs Magn\'etiques Intenses,
CNRS-UJF-UPS-INSA, 25, avenue des Martyrs, 38042 Grenoble, France}
\author{M. Orlita}\email{milan.orlita@lncmi.cnrs.fr}
\affiliation{Laboratoire National des Champs Magn\'etiques Intenses, CNRS-UJF-UPS-INSA, 25, avenue des
Martyrs, 38042 Grenoble, France}
\affiliation{Institute of Physics, Charles University, Ke Karlovu 5, CZ-121 16 Praha 2, Czech Republic}

\date{\today}

\begin{abstract}
The Faraday effect is a representative magneto-optical phenomenon, resulting from the
transfer of angular momentum between interacting light and matter in which
time-reversal symmetry has been broken by an externally applied magnetic field.
Here we report on the Faraday rotation induced in the prominent 3D topological insulator Bi$_2$Se$_3$ due to
bulk interband excitations. The origin of this non-resonant effect, extraordinarily
strong among other non-magnetic materials,
is traced back to the specific Dirac-type Hamiltonian for Bi$_2$Se$_3$, which
implies that electrons and holes in this material closely resemble relativistic
particles with a non-zero rest mass.
\end{abstract}


\maketitle

\vspace{0.2cm}

The recently emerged class of topological insulators (TIs)\cite{HsiehNature08,QiRMP11,HasanRMP10} comprises materials
with specific Dirac-type surface states, which continuously connect
otherwise well-separated conduction and valence bands. The existence of these intriguing
surface states is encoded in the specific bulk electronic band structures, which combines
band inversion with time-reversal symmetry. The sensitivity to perturbations breaking the time-reversal symmetry makes TIs, and in particular their surface states,
natural targets of Faraday rotation experiments~\cite{Valdes-AguilarPRL12,SushkovPRB10,JenkinsPRB13,WuCM15}, allowing us, among other things, to trace the
crossover from the topological to normal state of matter.
Opening the band gap in the TI surface states should be, at subgap photon energies, manifested by a Faraday angle determined only
by the fine structure constant $\alpha$~\cite{TsePRL10,MaciejkoPRL10}. Other universal Faraday rotation
effects have been proposed for Landau-quantized surface states of TIs~\cite{TsePRB10,TsePRB11}.
These become analogous to predictions for other quantum Hall systems, including graphene and electron/hole gases
in conventional 2D semiconductor heterostructures~\cite{MorimotoPRL09}, which have already been
tested in the very first experiments~\cite{IkebePRL10,CrasseeNP11,ShimanoNatureComm13}.

In this paper, we report on a strong Faraday rotation in the well-known Bi$_2$Se$_3$ 3D topological insulator. We show that the observed
effect appears due to interband excitations in bulk, from the valence to the partially filled conduction band. The strength of the
rotation, expressed in terms of the Verdet constant, is found to be extraordinary large for a non-magnetic material. This is related to the specific
bulk electronic band structure of Bi$_2$Se$_3$, which implies that charge carriers closely resemble massive relativistic particles,
with the spin-splitting large and equal for electrons and holes.

\begin{figure}[t]
      \includegraphics[width=8.5cm]{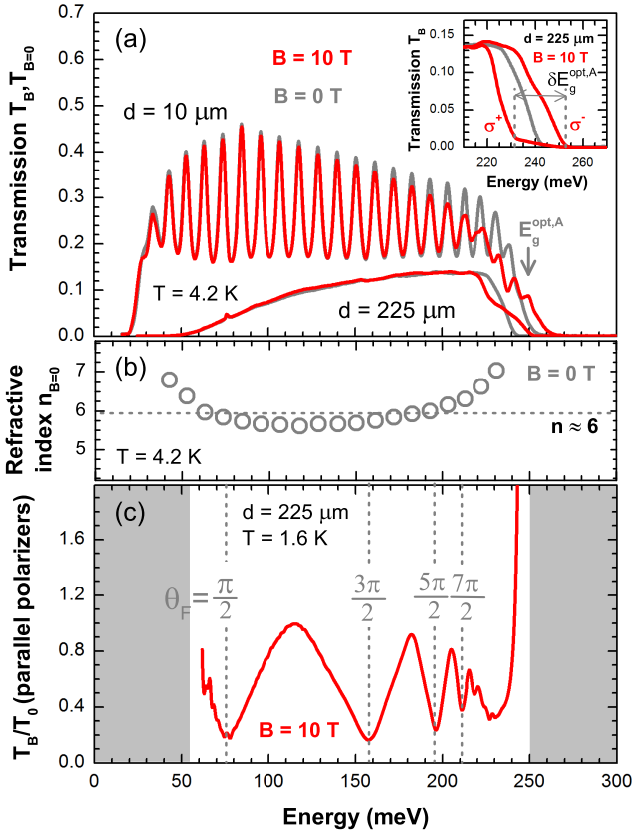}
      \caption{\label{SPKT} %(color online)
     (a): Low temperature infrared transmission of the sample A at B=0 and 10~T measured on free-standing layers
      with a thickness of 10 and 225~$\mu$m. The sample is transparent in the spectral window defined at lower energies by the plasma frequency $\hbar\omega_p$ and
      at higher energies by the interband absorption, implying the optical band gap of $E_g^{\mathrm{opt,A}}=E_g+E_F(1+m_e/m_h)\approx250$~meV due to the Burstein-Moss shift.
      This transparency window becomes narrower in thicker samples, due to free carrier absorption at low energies and due to the broadening of the interband absorption edge
      giving rise to non-zero absorption below $E_g^{\mathrm{opt,A}}$. When the magnetic field is applied, the interband absorption edge exhibits strong splitting $\delta E^{\mathrm{opt,A}}_g$, see Eq.~\ref{condition}, when probed by circularly
      polarized light, see the inset of the part (a). The pronounced modulation of the spectrum from the thinner sample are Fabry-P\'{e}rot interference fringes, which show high
      crystalline quality of the Bi$_2$Se$_3$ bulk specimen and provide us with an estimate of the refractive index plotted in the part (b). (c): Relative magneto-transmission of the 225-$\mu$m-thick
      free-standing Bi$_2$Se$_3$ layer prepared from sample A and placed in between two co-linear polarizers. The observed minima correspond to the Faraday
      rotation angle $(2k+1)\pi/2$ for $k=0,1,2\ldots$}
\end{figure}

\vspace{0.2cm}

\noindent\textbf{Results}

The Faraday rotation measurements have been performed on thin layers of Bi$_2$Se$_3$ sliced from bulk crystals (A, B and C) with various bulk electron densities
($N\approx1\times10^{18}$, $5\times10^{18}$ and $2\times10^{19}$~cm$^{-3}$). The prepared specimens with various thicknesses were characterized by infrared
transmission at $B=0$. The typical zero-field response observed is illustrated in Fig.~\ref{SPKT}a, where transmission spectra taken on two
free-standing layers prepared from the A crystal (with thicknesses of $d=10$ and 225~$\mu$m) are plotted. For the 10-$\mu$m-thick sample, the transmission window approximatively spans from the plasma frequency $\hbar\omega_p\approx20$~meV up to the interband absorption edge (optical band gap) $E_g^{\mathrm{opt,A}}=(250\pm10)$~meV. The difference between $E_g^{\mathrm{opt}}$ and the energy
band gap $E_g$, typical of doped degenerate semiconductors, is usually referred to as the Burstein-Moss shift~\cite{BursteinPR54}. For sample A,
this implies a zero-field Fermi level of $E_F^{A}\approx 30$~meV, assuming the parameters derived in Ref.~\onlinecite{OrlitaPRL15} ($m_e/m_h\approx0.8$ and $E_g\approx 200$~meV).

The pronounced Fabry-P\'{e}rot interference pattern observed in the transmission spectrum of the 10-$\mu$m-thick specimen, see Fig.~\ref{SPKT}a,
allows us to estimate the refractive index of Bi$_2$Se$_3$ at photon energies within the window of high transparency, as shown in Fig.~\ref{SPKT}b, with the averaged value
of $n\approx 6$. The absolute transmission of the 10-$\mu$m-thick layer, $T\approx0.3$, is close to the
theoretical value $T=2n/(n^2+1)\approx 2/n=1/3$ for non-absorbing medium characterized by the refractive index $n$. In the thicker sample,
the below-gap absorption is no longer negligible, implying significantly lower absolute transmission $T\approx0.1$ and also noticeably narrower transmission window.
The results obtained in transmission measurements on the other two specimens were analogous, providing us with the optical band gaps of
$E_g^{\mathrm{opt,B}}=(340\pm10)$~meV and $E_g^{\mathrm{opt,C}}=(400\pm10)$~meV. The declared errors here are mostly due to the variation of the electron density across the bulk crystals.

Interestingly, the application of the magnetic field gives rise to a strong modification of the interband absorption edge of Bi$_2$Se$_3$. Namely, a splitting
with respect to the circular polarization of the probing radiation appears, as shown in the inset of Fig.~\ref{SPKT}a. This splitting was found to be linear in $B$
and the same for all the three investigated crystals: $\delta E^{\mathrm{opt}}_g\approx 2.3$~meV/T. In the transmission spectrum taken with non-polarized (or linearly polarized) radiation,
this splitting manifests itself as a characteristic step-like profile of the interband absorption edge, see Fig.~\ref{SPKT}a. This significant difference in interband
absorption for circularly polarized light of the opposite helicity is the origin of the strong interband Faraday rotation discussed in this paper.

\begin{figure}[t]
      \includegraphics[width=8.5cm]{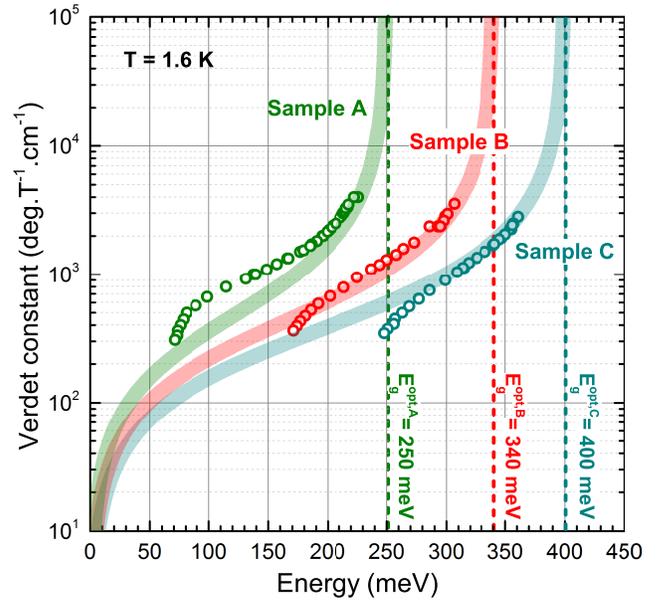}
      \caption{\label{Fig2} %(color online)
      The experimentally determined Verdet constant for samples A, B and C as a function of the photon energy. The theoretical curves have been plotted
      using Eg.~\ref{VerdetBi2Se3}, their widths reflect the uncertainty in the determination of the interband absorption edges and sample thicknesses.}
\end{figure}

The extraordinarily strong Faraday effect can be probed using a simple configuration with the sample placed in between two co-linearly oriented polarizers.
The Faraday effect is then manifested by a characteristic modulation of the relative magneto-transmission spectra $T_B/T_0$, see Fig.~\ref{SPKT}c. This spectrum
has been taken at $B=10$~T on the 225-$\mu$m-thick layer prepared from the crystal A. The pronounced minima can be easily identified with particular Faraday angles $\theta_F=\frac{\pi}{2}(2k+1)$, $k=0,1,2\ldots$, and
the corresponding Verdet constants (normalized Faraday angles), $V(\omega)=\theta_F/(B.d)$, can be calculated from $T_B/T_0$ curves measured at various values of $B$.
The frequency dependence of the Verdet constant, deduced for samples A, B and C, has been plotted in Fig.~\ref{Fig2}. Notably, no field-dependence
of the Verdet constant has been revealed within the range of the magnetic field applied (up to 13~T).

\vspace{0.2cm}

\noindent\textbf{Discussion}

\begin{figure}[t]
      \includegraphics[width=8.5cm]{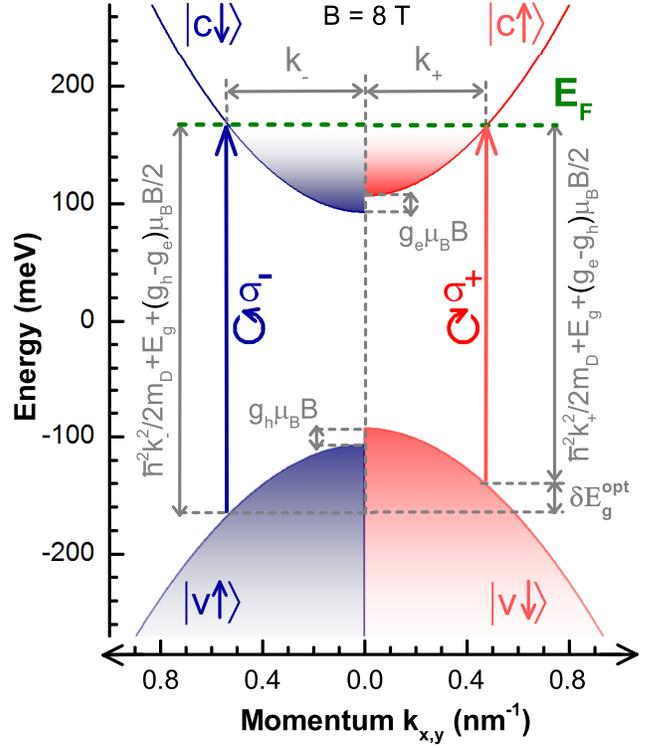}
      \caption{\label{BandStructure} %(color online)
      Nearly parabolic conduction and valence bands of Bi$_2$Se$_3$,
      around the $\Gamma$ point of the Brillouin zone, with a significant spin-splitting due to an externally applied
      magnetic field ($B=8$~T). The vertical arrows denote the lowest energy interband absorption in right and left
      polarized radiation, split in energy by $\delta E_g^{\mathrm{opt}}=\mu_B B [g_e(m_e/m_h)+g_h]$. This splitting may be easily derived
       when we consider that the lowest momenta $k_+$ and $k_-$, for which interband absorption is allowed in a given circular polarization,
       satisfy the condition: $\hbar^2 (k_-^2-k^2_+)/(2m_e)=g_e \mu_B B$.
      $m_D$ used in formulae in the figure stands for the reduced mass $1/m_D=1/m_e+1/m_h$, which for Bi$_2$Se$_3$ equals to the Dirac mass $m_D=E_g/(2v_D^2)$.}
\end{figure}

To account for the observed magneto-optical response let us recall the band structure of Bi$_2$Se$_3$. Within the past few years, the exact shape of electronic
bands in this material has been subject of vast discussions. Nevertheless, Bi$_2$Se$_3$ is most likely a direct band-gap semiconductor, see, \textit{e.g.}, Refs.~\onlinecite{YazyevPRB12,AguileraPRB13}, which can be well-described by Dirac-type models such as the one proposed in Refs.~\onlinecite{ZhangNaturePhys09,LiuPRB10}. Our recent
magneto-optical study~\cite{OrlitaPRL15} implies that the conduction and valence bands are nearly parabolic with a high degree of the electron-hole symmetry, see Fig.~\ref{BandStructure}.
This electronic band structure can be described using a simplified Dirac-type Hamiltonian for massive particles, which
contains only two parameters: the band gap $E_g$ and the velocity parameter $v_D$.
These two parameters provide us with reasonably accurate estimates for the effective masses and $g$ factors: $m_e\approx m_h \approx 2m_D$ and $g_e\approx g_h\approx 2m_0/m_D$,
respectively, where $m_D=E_g/(2v_D^2)$ is the Dirac mass.

Large $g$ factors in Bi$_2$Se$_3$ give rise to a specific regime at low magnetic fields, when a pronounced spin-splitting of electronic states appears,
but Landau levels are still not well resolved ($\mu.B<1$), as schematically sketched in Fig.~\ref{BandStructure}. Within this picture of spin-split bands,
the interband absorption edge splits with respect to the circular polarization of light:
\begin{equation}
\label{condition}
\delta E_g^{\mathrm{opt}}=\mu_B B \left(g_e\frac{m_e}{m_h}+g_h\right),
\end{equation}
where $\mu_B$ is the Bohr magneton.
This formula may be straightforwardly derived, using purely geometrical arguments, see Fig.~\ref{BandStructure} and its caption.

Notably, due to nearly equal $g$ factors of electrons and holes in Bi$_2$Se$_3$ ($g_e\approx g_h$), a pronounced splitting of the interband
absorption edge \eqref{condition} only appears in this material when the conduction band is partially filled by electrons (or alternatively the
valence band by holes). The observed Faraday rotation, even though primarily related to the circular dichroism of the interband absorption, is thus
basically induced by the presence of free conduction-band electrons. However, this effect should be clearly distinguished from the intraband
Faraday rotation due to free charge carriers, where the circular dichroism originates in cyclotron motion of particles
and related resonant absorption\cite{BennettPR65}.

Importantly, the splitting \eqref{condition} does not explicitly depend on the electron density, in agreement with our experiments.
The carrier density only determines the saturation field $B_s$, at which a
full spin-polarization of electrons is achieved, $B_s=2^{2/3} E_F /(\mu_B g_e)$~\cite{MukhopadhyayPRB15}. Above $B_s$, the formula \eqref{condition}
is no longer valid and the splitting saturates at $\delta E_g^{\mathrm{opt}} \approx 2^{2/3} E_F (1+m_e/m_h)$ when $g_e\approx g_h$.
Taking the parameters derived in Ref.~\onlinecite{OrlitaPRL15}
($m_e/m_h\approx 0.8$, $g_e \approx 27$ and $g_h\approx 24$) we get $\delta E_g^{\mathrm{opt}}/B \approx 2.6$~meV/T in very good agreement
with the experiment. The saturation field for the lowest doped specimen A should reach $B_s\approx 20$~T, well above the
magnetic fields applied in the experiments presented here.

The Verdet constant is proportional to the difference between refractive indices
for right ``+'' and left ``-'' circularly polarized light: $V(\omega)=\omega(n^+-n^-)/(2cB)$, which can be approximated in a weakly
absorbing medium  by~\cite{BoswarvaPRS62}:
\begin{equation}
\label{Verdet}
V(\omega) \approx \frac{\omega}{4 c B \bar{n}}\left(\varepsilon_{1}^{+}-\varepsilon_{1}^{-}\right),
\end{equation}
where $\varepsilon_{1}^{\pm}$ stands for the real part of the dielectric function, $\bar{n}=(n^++n^-)/2\approx n$ and $c$ is the speed of light in vacuum.

When we neglect the contribution from all electronic bands other than the conduction and valence bands and
assume the interband matrix elements to be independent of the magnetic field and momentum, $|P^\pm_{cv}(\mathbf{k})|^2\approx2m_0^2v_D^2$~\cite{LiuPRB10},
the imaginary (dissipative) part of the dielectric function at low magnetic fields and temperatures reads, for a given circular polarization~\cite{CardonaYu}:
\begin{equation}
\label{Cardona}
\varepsilon^\pm_{2} (\omega)= \frac{2\pi}{\varepsilon_{0}}\frac{e^{2}v_D^2}{\omega^2}j^{\pm}(\omega)\,\,\Theta(\hbar\omega-E_g^{\mathrm{opt}}\mp\delta E_g^{\mathrm{opt}}/2),
\end{equation}
where $j^{\pm}(\omega)$ is the joint density of states and $\Theta$ is the Heaviside step function, which describes
the low-energy onset of interband absorption at the photon energy of $E_g^{\mathrm{opt}}=E_g+E_F(1+m_e/m_h)$.

Assuming strictly equal spin splitting for electrons and holes ($g_e=g_h$) and neglecting the anisotropy of electronic bands,
the joint density of states becomes identical for both circular polarizations, $j^{\pm}(\omega) = \sqrt{\hbar\omega-E_g}m_D^{3/2}/(\sqrt{2}\pi^{2}\hbar^3)$,
and the imaginary part of the dielectric function \eqref{Cardona} takes the form:
\begin{equation}
\varepsilon_{2}^{\pm} (\omega)= \frac{A}{\omega^2} \sqrt{\hbar\omega-E_g} \,\, \Theta(\hbar\omega-E_g^{\mathrm{opt}}\mp\delta E_g^{\mathrm{opt}}/2),
\end{equation}
where $A = \sqrt{2} e^2 v_D^2 m_D^{3/2}/(\varepsilon_{0}\pi\hbar^{3})$.

Using the Kramers-Kronig relations applied to $\varepsilon^{\pm}_2(\omega)$ together with Eq.~\ref{Verdet}, the
Verdet constant can be expressed as:
\begin{equation}
\label{Kramers}
V(\omega)= \frac{A\omega \hbar^2}{B 2\pi c n} \int_{E_g^{\mathrm{opt}}-\delta E_g^{\mathrm{opt}}/2}^{E_g^{\mathrm{opt}}+\delta E_g^{\mathrm{opt}}/2}\frac{\sqrt{\xi-E_g}}{\xi(\xi^{2}-\hbar^2\omega^{2})}d\xi.
\end{equation}
Taking assumption of the full electron-hole symmetry ($m_e=m_h$),
we finally get, in the limit of low magnetic fields ($\delta E_g^{\mathrm{opt}} \ll E_F$), an approximative expression:
($\alpha$ is the fine structure constant):
\begin{equation}
\label{VerdetBi2Se3}
%V(\omega)= \frac{8e\alpha}{\pi n}\frac{\hbar\omega}{E_g}\frac{[E_F E_g]^{1/2}}{E_g^2-\hbar^2\omega^2}v_D\newline
V(\omega)= v_D \frac{8e\alpha}{\pi n}\frac{\sqrt{E_F E_g}}{E_g^{\mathrm{opt}}}\frac{\hbar\omega}{(E_g^{\mathrm{opt}})^2-\hbar^2\omega^2}.
\end{equation}

It is worth noting that in the low-field limit, the density of spin-polarized
carriers,  $\delta N= N^{\downarrow}-N^{\uparrow}$, can be approximated as the spin-splitting, $E_s=g_e \mu B$,
multiplied by the zero-field density of states, $DoS(E_F)\propto \sqrt{E_F}$. The Faraday angle corresponding to Eq.~\ref{VerdetBi2Se3},
$\theta_F = V(\omega) B d \propto B.\sqrt{E_F}$, thus becomes directly proportional to  $\delta N$. This may resemble the interband Faraday effect due to the Raman spin-flip
of electrons bound to In-donors in CdS, reported by Romestain~\emph{et~al.}~\cite{RomestainPRL75}, and later on,
of free conduction-band electrons in InSb and HgCdTe~\cite{YuenJVST87,AggarwalAPL88}. However, the mechanism
of the Faraday rotation reported here is different from that considered in Refs.~\onlinecite{RomestainPRL75,YuenJVST87,AggarwalAPL88}
and is uniquely related to dipole-allowed interband absorption.
While in the Raman spin-flip induced Faraday rotation, the angle becomes proportional to the spin splitting of electrons ($g_e$),
in our case, the resulting rotation is clearly sensitive to the spin splitting
of both electrons and holes (both $g_e$ and $g_h$ factors), as seen from the initial Eq.~\ref{condition}.

Scaling the photon energy with respect to the optical band gap, $x=\hbar\omega/E_g^{\mathrm{opt}}$, we get the characteristic frequency profile of the interband
Faraday rotation: $V(x)\propto x/(1-x^2)$, $0\leq x<1$. Interestingly, this profile has a considerably simpler form as compared to the
expressions derived for and applied to undoped semiconductors~\cite{BoswarvaPRS62,RothPR64,BalkanskiJPCS66,BennettPR65}. This is due to the
match between the electron and hole spin splitting in Bi$_2$Se$_3$, $g_e=g_h$, which implies the same joint density of states for both circular polarizations,
and therefore, only the finite integration range in the Kramers-Kronig transformation, see Eq.~\ref{Kramers}. Another simplification,
which again comes directly from the Dirac-type Hamiltonian valid for Bi$_2$Se$_3$, is rather high degree of electron-hole symmetry ($m_e\approx m_h$). This
allows us to express the optical band gap as $E_g^{\mathrm{opt}}-E_g\approx 2E_F$.

We should stress that the formula \eqref{VerdetBi2Se3} derived for the Verdet constant in Bi$_2$Se$_3$ does not contain any tunable parameters.
The bulk band gap $E_g=200$~meV and the velocity parameter $v_D=0.47\times10^6$~m/s are known from experiments, for instance, from our recent
Landau level spectroscopy of thin Bi$_2$Se$_3$ layers~\cite{OrlitaPRL15}. The refractive index $n\approx 6$ and
the optical band gaps $E_g^{\mathrm{opt}}$ are directly read from transmission spectra of thin specimens at $B=0$, as
illustrated for sample A in Figs.~\ref{SPKT}a and b.

The theoretical curves, calculated using these parameters in Eq.~\ref{VerdetBi2Se3}, are in good quantitative agreement with the experimentally determined
Verdet constant, see Fig.~\ref{Fig2}. For the crystals B and C, the major deviations
appear only at lower energies, where the contribution to the total Faraday rotation arising from the
cyclotron resonance absorption~\cite{BennettPR65} (fully intraband process) is no longer negligible.
For the lowest-doped sample A, the spin-splitting becomes, for the studied range of magnetic fields,
comparable with the Fermi level, which brings the formula \eqref{VerdetBi2Se3} close to the limit
of its validity.

Let us now compare our results with the Faraday rotation on other materials,
the choice of which remains, in the equivalent spectral range, still limited. Taking the reference at the wavelength 10.6~$\mu$m
of the CO$_2$ laser, the highest specific interband Faraday rotation for non-magnetic materials has been reported for InSb,
$V_{\mathrm{InSb}} = 10-100$~deg/(T.cm)~\cite{SmithPIJPS62,JimenezPRB94}. Interestingly, this semiconductor has a band gap nearly
identical to Bi$_2$Se$_3$. This rotation is at least one order of magnitude lower
compared to the value we have found in Bi$_2$Se$_3$: $V_{\mathrm{Bi_2Se_3}} \approx 10^3$~deg/(T.cm), see the Verdet constant
at wavelength of 10.6~$\mu$m (corresponding to the photon energy of $\approx$120~meV) for the lowest doped specimen A in Fig.~\ref{Fig2}.
In fact, the observed Verdet constants become comparable to values known for magnetic semiconductors, see, \emph{e.g.}, Ref.~\onlinecite{GajSSC78}.

The relatively high Verdet constants of Bi$_2$Se$_3$ invoke the possibility to use
it as the active medium in Faraday rotators/insulators although
one has to keep in mind the Drude-type absorption on free conduction-band electrons, which lowers the overall transmission.
This absorption may be reduced, by decreasing the electron density, nevertheless, this may be a challenging task for the
current growth technology of this material. Moreover, this would also lower the saturation field $B_s$.

It is instructive to discuss the implications of our results on other 3D TIs from the same family, such as
Bi$_2$Te$_3$ or Sb$_2$Te$_3$. Most likely, their bands strongly deviate from the parabolic profiles~\cite{MichiardiPRB14},
which are characteristic of Bi$_2$Se$_3$, but still they are characterized by similar band gaps, and also, as seen, for instance,
from ARPES experiments~\cite{HsiehPRL09}, similar velocity parameters. Since these materials are described by the same
expanded 3D Dirac Hamiltonian~\cite{LiuPRB10,ZhangNaturePhys09}, we may expect the electron and hole $g$ factors
to roughly follow the simple estimate $g_e\approx g_h\approx 4m_0v_D^2/E_g$ deduced for Bi$_2$Se$_3$ in Ref.~\onlinecite{OrlitaPRL15}.
This would imply an analogous splitting of the interband absorption edge given by Eq.~\ref{condition} and equally strong interband Faraday effect in doped TIs.

To conclude, we have probed the Faraday rotation induced by interband excitations in a series
of bulk Bi$_2$Se$_3$ specimens. We show that this effect is at least by an order of magnitude
stronger than in other non-magnetic materials. We demonstrate that the particular strength of the effect has its origin in
the relativistic-like Hamiltonian applicable to Bi$_2$Se$_3$ thanks to which
electrons and holes behave as massive Dirac particles. A simple formula based on this two-parameter Dirac-type Hamiltonian
is derived to describe this phenomenon quantitatively, requiring no tunable parameters.
We also predict that similarly strong interband Faraday effect should be present in other 3D topological insulators, in particular in those from
the Bi$_2$Se$_3$ family.

\vspace{0.4cm}

{\small \textbf{Methods:}
The studied Bi$_2$Se$_3$ crystals have been prepared using the standard Bridgman method. The starting material for growing the single crystals was prepared from the elements Bi and Se of 5N purity. Polycrystalline material was prepared from a mixture of the elements close to stoichiometry (Bi:Se = 2:3) in silica ampoules evacuated to a pressure of 10$^{-4}$~Pa. The synthesis was carried out at the temperature of 1073~K. A conical quartz ampoule, containing the synthesized polycrystalline material, was then placed in the upper (warmer) part of the Bridgman furnace, where it was remelted. Then it was lowered into a temperature gradient of 80 K/cm (30 K/cm for the sample C) at a rate of 1.3 mm/h. Three bulk crystals, differing in the concentration of conduction-band electrons in the conduction band, have been chosen for this study. They were characterized by approximate electron densities $N\approx1\times10^{18}$, $5\times10^{18}$ and $2\times10^{19}$~cm$^{-3}$ and denoted as samples A, B and C, respectively.

The prepared single crystals, easily cleavable along the hexagonal planes (0001), were sliced, using a microtome machine to free-standing
layers with various thicknesses. All experiments, magneto-transmission and Faraday-angle measurements, were performed in the Faraday
configuration with light propagating along the $c$ axis of Bi$_2$Se$_3$. A macroscopic area of
the sample ($\sim$4~mm$^2$) was exposed to the radiation of a globar, which was analysed by a Fourier transform spectrometer
and, using light-pipe optics, delivered to the sample placed in a superconducting magnet. The transmitted light was
detected by a composite bolometer placed directly below the sample, kept at a temperature of 1.6~or 4.2~K. To measure the Faraday rotation, the specimens were
placed in between two co-linearly oriented wire-grid polarizers defined holographically on a KRS-5 substrate.  In experiments performed
with circularly polarized light, a linear polarizer and a zero-order MgF$_2$ quarter wave plate (centered at $\lambda = 5$~$\mu$m) were used.

\vspace{0.2cm}

\textbf{Authors contributions:}
M.O., G.M. and M.P. conceived the study. Experiments were performed by L.O., M.H., G.M., C.F., B.A.P. and M.O.
The samples were prepared and characterized by C.D., A.M., G.S., A.H., M.V.Y. and R.W.M.
The theoretical model was developed by M.V., M.P. and M.O. The manuscript was written by M.V. and M.O.
All authors discussed the results and commented on the manuscript.

\vspace{0.4cm}

\textbf{Acknowledgments:}
The work has been supported by the ERC project MOMB. Authors
acknowledge discussions with D. M. Basko, S. A. Studenikin, T. Brauner, C. Michel, E. M. Hankiewicz and M. O. Goerbig. The work in ITME-Poland has been supported
by the research project NCN UMO, 2011/03/B/ST3/03362 Polska. M.V.Y acknowledges the support from RFBR though projects 14-02-00080, 14-03-00121 and UB RAS 15-20-3-11.

\vspace{0.2cm}

\textbf{Competing financial interests:}
The authors declare no competing financial interests.}

%\bibliography{Bi2Se3}

%

\end{document}